\documentclass[conference]{IEEEtran}
\usepackage{tabularx}
\usepackage{multirow}
\usepackage{graphicx}

\begin{document}






%

\title{Latency Evaluation of a Virtualized MME}
\author{\IEEEauthorblockN{Jonathan Prados-Garzon, Juan J. Ramos-Munoz, Pablo Ameigeiras, Pilar Andres-Maldonado, Juan M. Lopez-Soler}
\IEEEauthorblockA{Research Center on Information and Communications Technologies\\
University of Granada\\
Granada, Spain\\
Emails: jpg@ugr.es, jjramos@ugr.es, pameigeiras@ugr.es, pam91@correo.ugr.es, juanma@ugr.es}}
\maketitle

\date{20 September 2015}

\maketitle
\begin{abstract}

 Network Virtualization is one of the key technologies for developing the future mobile networks. However, the performance of virtual mobile entities may not be sufficient for delivering the service required for future networks in terms of throughput or service time. In addition, to take advantage of the virtualization capabilities, a criterion to decide when to scale out the number of instances is a must. 
  In this paper we propose an LTE \emph{virtualized Mobility Management Entity} queue model to evaluate its service time for a given signaling workload. The estimation of this latency can serve to decide how many processing instances should be deployed to provide a target service. Additionally, we provide a compound data traffic model for the future mobile applications, and we predict theoretically the control workload that it will generate. Finally, we evaluate the \emph{virtualized Mobility Management Entity} overall delay by simulation, providing insights for selecting the number of virtual instances for a given number of users.  
\end{abstract}

%
%


%
%

%
%

\begin{IEEEkeywords}
virtualized MME, queue model, NFV, LTE.
\end{IEEEkeywords}

\section{Introduction}
 
 Nowadays, telecom industry is regarding Network Virtualization as one of the key technologies in the future cellular networks. Network Functions Virtualization (NFV) offers the operators the possibility of running the network functions on industry standard high volume servers instead of using expensive and vendor-dependent hardware \cite{nfv-etsi}\cite{nfv-survey}. The decomposition of a service in a set of Virtual Network Functions (VNF) which can be executed in standard servers, allows for instantiating these VNFs in different network locations as needed. 
 Concretely, NFV promises to enable organizations to: i) reduce capital and operational expenditures, ii) accelerate time-to-market of new services, iii) deliver agility and flexibility, and iv) scale up services on demand \cite{nfv-etsi}. 

  However, it is unclear whether virtualized entities will be able to cope with the demanding requirements that future mobile networks will have to face, such as tight service latency deadlines or very high data and signaling traffic rates. 
  Some works have addressed the study of the feasibility of the virtualization of the LTE Evolved Packet Core (LTE/EPC). 
  For instance, the authors of \cite{Hirschman15} implement an entire EPC in general purpose processors. They show that servicing the synthetic workload generated by 50000 users is viable. However, they do not analyze the impact of scaling the resources on the overall service time. 
   In \cite{rajan15}, the authors point out potential bottlenecks of  a virtualized EPC (vEPC). To that end, they propose a simple queue model to estimate the service time of their vEPC implementation. Experimentally, they show that the \emph{Serving Gateway} (S-GW) may represent the bottleneck of a vEPC, and demonstrate that the control plane signaling may interfere with the user plane packet processing. They conclude that the direct implementation of the EPC entities in virtualized servers degrades the system performance, requiring thus a thorough new design.
   
  The present work focuses on estimating how the signaling plane workloads expected for the near future affects the service time of a \emph{virtualized Mobility Management Entity} (vMME) which can scale its resources. This is the first step to predict the resources needed to provide low latency services.
  The contribution of this paper is threefold. First, we propose a detailed queue model of a Virtual MME in a datacenter. To do that, we calculate experimentally the service rates of the vMME processes. Second, we characterize theoretically and by simulation the control messages rate generated by the users's activity. 
  Third, we characterize the service time of a vMME for different control plane workloads. As a result, we provide the estimation of the system delay depending on the number of network users and vMME instances.

 The paper is organized as follows. Section \ref{sec:system-model} describes the architecture of the vMME analyzed. 
In sections \ref{sec:traffic-models} and \ref{sec:arrival-rate}, the user and control planes traffics are modeled. In section \ref{sec:queuing-model} we present a queue system model for the vMME, which is simulated and evaluated in section \ref{sec:numerical-results}. Finally, section \ref{sec:conclusions} draws the main conclusions of this study.
 

 
 
 
 

\section{System Model}
\label{sec:system-model}

\begin{figure}[tb]
\begin{center}
\includegraphics[width=0.8\columnwidth]{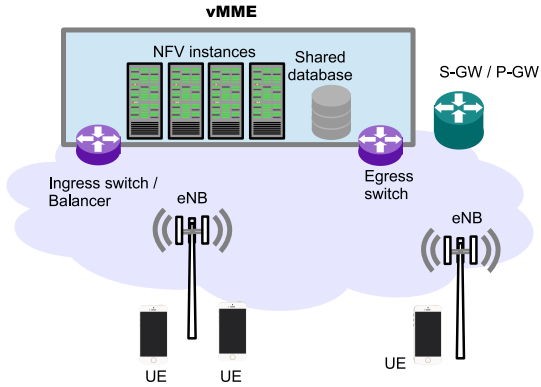}
\end{center}
\caption{Overall system model.}
\label{fig:5g-arch}
\end{figure}

 In this work, we assume a general LTE/EPC network architecture (see Figure \ref{fig:5g-arch}), with a logically centralized vMME, which runs in a cloud computing facility. For simplicity, we will assume that every processor in the data center provides the same computational power.

\subsection{System Architecture}

 

 
 
 For the LTE/EPC considered in this work, the principal system entities are:
\begin{itemize}
	\item The \emph{User Equipment (UE)}. The UE represents the terminals which allow each user to connect to the network via the eNodeB base stations. The UEs run the users' applications which generate or consume network traffic. 
    The activity of the UE and the generation of network traffic trigger the network control procedures.
  \item The virtualized \emph{Mobility Management Entity (vMME)}. This is the main control entity, and the responsible of maintaining the mobility state of the UE, and the management of bearers and QoS provision of flows from and to the UE. In this architecture, the vMME is implemented as NVF instances in the virtualization facility. This allows to deploy several MME instances to scale with the network growth, sharing the processing between the active MME instances. To do this, protocol and users' state are stored in a shared database.
  To ease the processing sharing, the datacenter \emph{ingress switch} acts as a balancer, scheduling each process request to the MME instance with the lowest processing load. To increase the scalability, resilience and performance of the NFV processing, each procedure is split into request and response transactions. When an NFV finishes its processing, it saves the transaction state into the shared database. When a subsequent request arrives to an MME instance, it first gathers the transaction state from the database to continue from. 
  This way, if a procedure comprises several stages which depend on the completion of the precedent stage, the NVF function will not be blocked. 

 \end{itemize}

\subsection{Control Plane Procedures} 
\label{sec:control-plane-procedures}

There exist several signaling procedures in LTE that allow the control plane to manage the UE mobility and the data flow between the UE and \emph{Packet Data Network Gateway} (PDN-GW). From all of them, we only concentrate on the ones that generate most signaling load \cite{Hirschman15}. Each procedure typically implies an exchange of signaling messages between the control plane entities \cite{lte-signaling}. When the MME receives one of these messages, it processes the message, and later it possibly sends a new message to the another entity. In the following subsections, we describe the processing carried out by the MME during the main control plane procedures \cite{lte-signaling}.

 \subsubsection{Service Request (SR)}
 
When a UE does not have  available resources and new traffic is generated, either from this UE, or from the network to this UE, the UE starts a Service Request (SR) procedure. We focus on the UE-triggered SR. During this procedure the MME receives three different messages: an Initial UE Message ($SR_1$), an Initial Context Setup Response ($SR_2$), and a Modify Bearer Response ($SR_3$).


To process the Initial UE Message ($SR_1$) the MME has to carry out UE integrity check and message decrypting. Additionally, it generates identifiers for the bearers to be established. Moreover, it stores and retrieves parameters and variables related to the UE context. Some of them are included in the subsequent Initial Context Setup Request message. During the processing of the Initial Context Setup Response message ($SR_2$), the MME also retrieves information of the UE context, and includes this information in the subsequent Modify Bearer Request message. The processing of the Modify Bearer Response ($SR_3$) is minimum as this message is only a confirmation.

 \subsubsection{Service Release (SRR)}
 
The Service Release (SRR) procedure is triggered by user inactivity. Its purpose is to release a data plane bearer and its control plane connection for a UE. During the SRR, the MME processes three messages: a UE Context Release Request ($SRR_1$), a Release Access Bearers Response ($SRR_2$), and a UE Context Release Complete ($SRR_3$).
  

To process both the UE Context Release Request message ($SRR_1$) and the Release Access Bearers Request ($SRR_2$), the MME needs to retrieve information of the UE context, and include this information in the subsequent messages. The processing of the UE Context Release Complete message ($SRR_3$) mainly implies the deletion of the bearer's context information by the MME.

 \subsubsection{X2-Based Handover}
  
The MME participates in the X2-based Handover (HO) during the handover completion phase. Its purpose is to switch the bearers end point from the source to the target eNB. The MME receives two messages during this phase: a Path Switch Request message ($HR_1$) and a  Modify Bearer Response ($HR_2$).


To process both the Path Switch Request message ($HR_1$) and the Modify Bearer Response ($HR_2$), the MME also needs to retrieve information of the UE context, and include this information in the subsequent messages. To process the Path Switch Request message, the MME also needs to store new information such as the ids of the new serving cell and new tracking area.

 


 
\section{Application Traffic Models}
\label{sec:traffic-models}

\begin{table*}[tb]
\centering
\caption{Traffic models characterization}
\label{tab:traffic-models}
{\small
\begin{tabularx}{\textwidth}{l|l|X}
\hline
\hline
Traffic Type	& Parameters	& Statistical  Characterization \\                                                      \hline
\hline
\multirow{6}{*}{\parbox[t]{1.7cm}{\centering Web \\browsing \\(HTTP) $P_{app}=0.74$}}	& Main Object Size	& Truncated Lognormal Distribution: $\mu$=15.098 $\sigma$=4.390E-5 min=100 Bytes max=6 MBytes 	\\  \cline{2-3}                                                                            
                                        & Embedded Object Size	& Truncated Lognormal Distribution: $\mu$=6.17 $\sigma$=2.36 min=50 Bytes max=2 MBytes
\\ \cline{2-3}
                                        & Number of Embedded Objects per Page & Truncated Pareto Distribution: mean=22 shape=1.1
\\ \cline{2-3}
                                        & Parsing Time                        & Exponential Distribution: mean=0.13 seconds
\\ \cline{2-3}
                                        & Reading Time                        & Exponential Distribution: mean=30 seconds
                                        \\ \cline{2-3}
                                        & Number of pageviews per session     & Geometric Distribution: p=0.893 mean=9.312
                                        \\ \cline{2-3}
\hline
\multirow{4}{*}{\parbox[t]{1.7cm}{\centering HTTP \\progressive \\video $P_{app}=0.03$}} & Video Encoding Rate                 & Uniform distribution with ranges:  $(2.5, 3.0)$ Mbps / (4.0,4.5) Mbps / $(12.5, 16.0)$ Mbps / $(20.0,25.0)$ Mbps, for equiprobable itags: 137 / 264 / 266 / 315 respectively. \\ \cline{2-3}
                                        & Video Duration                      & Distribution extracted from \cite{Ameigeiras12}                                                                                                                                                                                                  \\ \cline{2-3}
                                        & Reading Time                    & Exponential Distribution: mean=30 seconds                                                                                                                                                                        \\ \cline{2-3}
                                        & Number of videoviews per session    & Geometric Distribution: p=0.6 mean=2.5                                                                                                                                                                          \\ \cline{2-3}
\hline
\multirow{2}{*}{\parbox[t]{1.9cm}{\centering Video calling $P_{app}=0.23$}}          & Call Holding Time                   & Pareto Distribution: k=-0.39 s=69.33 m=0                                                                                                                                                                        \\ \cline{2-3}
                                        & Number of calls per session         & Constant = 1
\\ \cline{2-3}                                       
\hline \hline                                                  
\end{tabularx}
}
\end{table*}

This section describes the application models considered in this work along with their statistical characterization. 

Applications generate traffic during the \emph{application activity periods} of a \emph{session}. A \emph{session} is the UE activity comprised between the instant the user launches a network application and the time the user closes or stops using it. The \emph{application activity periods} are time intervals in which the application is transmitting or receiving data. 

A session consists of $N$ application activity periods ($T_{on}$) separated by $N-1$ \emph{reading times}. The \emph{reading time} ($D$) is the time period between two successive activity periods. During the reading time the user does actions such as reading the downloaded webpage or deciding the next video to watch. The time between the start of two consecutive sessions is the \emph{inter arrival session time} $IAST$. Based on \cite{Ilias14}, we configure $IAST$ to follow an exponential distribution with a mean of 1200 seconds.

When a session begins, the user chooses a certain application with a given probability $P_{app}$ (see Table \ref{tab:traffic-models}). Three types of applications are considered in this work: i) web browsing, ii) HTTP progressive video and iii) video calling. 
To provide the data rates of the future mobile traffic, we have followed the predictions assumed in the METIS project \cite{metis}. The statistical characterization of these application models are summarized in Table \ref{tab:traffic-models} and described in the following subsections.

\subsection{Web Browsing}

 The characterization of this traffic is described in \cite{ngmn08}. The amount of data downloaded for an application activity period (i.e., webpage size) of a web browsing session is determined by the main object size (i.e. the HTML file), the number of embedded objects and their sizes. During a session, the number of downloaded webpages per session is set to follow a geometric distribution which fits the data of \cite{Gou11}.

 The download time is determined by the webpage size, the link data rate, and the \emph{parsing time}. The \emph{parsing time} is the time  the web browser takes to parse the embedded objects. 

 We have set the future webpages sizes by extrapolating the data series of \cite{httpArchive}, and scaling main objects size accordingly. 
 
\subsection{HTTP Progressive Video}

 This application model follows the YouTube operation described in \cite{Ameigeiras12}, in which a video is transferred at a constant limited rate during a \emph{throttling phase} after an initial period of high downloading rate, called \emph{initial burst}. The number of downloaded video clips per session is set to follow a geometric distribution which fits the data of \cite{Phillipa08}.  The \emph{reading time} is assumed to be similar to the one of the web browsing case (see table \ref{tab:traffic-models}). 
 
 The size of each video is calculated from its duration and encoding rate. The video encoding rate depends on the video format selected. Each video format, identified by an \emph{itag} number, determines a container file format, an encoding algorithm, and a video resolution. To meet the METIS predicted data rates, we have considered the YouTube video formats with the highest encoding rates and resolutions.  
 
 The video download time (i.e., activity period) is determined by the bottleneck link data rate during the initial burst, and limited by the media server during the throttling phase \cite{Ameigeiras12}.
 

\subsection{Video calling}
 For this application, a session starts when the user opens a video calling client and makes a single call to someone. This application generates constant bit rate traffic of 1.5 Mbps which is the recommended download/upload speed of Skype for HD video calling.
 
 The call duration or \emph{call holding time} determines the application activity period duration. The statistical characterization for the call duration has been extracted from \cite{Trang04}.


\section{Signaling procedures rate characterization}
\label{sec:arrival-rate}

 User's activity may trigger network control procedures which the vMME has to process. The frequency of these requests affects the vMME performance. In this section we derive mathematical expressions to predict the rate of procedure requests, which depend on the user's activity. 

An SR procedure occurs whenever a user application is going to start an activity period without network resources assigned. 
When an application activity period finishes, a \emph{user inactivity timer} with a value of $T_{I}$ starts. If this timer expires before the user application starts a new activity period, the SRR procedure is triggered. 

The mean SR arrival rate per user $\lambda_{U}^{SR}$ is defined as the average number of SR procedures triggered by a user per unit time. Hence, the $\lambda_{U}^{SR}$ can be computed by multiplying the mean number of SRs procedures per session by the mean session arrival rate $\lambda_{S}$. In turn, the mean number of SRs per session is the mean number of application activity periods per session $\overline{N}$ times the probability that the inactivity timer expires. The first activity period begins after an inter session time $T_{IS}$ (i.e. the time elapsed from the end of a session to the beginning of the next one), while the following $N-1$ activity periods begin after each reading time. Thus, $\lambda_{U}^{SR}$ can be calculated as: 

\begin{equation}{\lambda_{U}^{SR}=\lambda_{S}\cdot ((\overline{N}-1)\cdot P(D>T_{I}) + P(T_{IS}>T_{I}))}
\label{eq:lambdaSR}
\end{equation}

 Since each SR have a corresponding SRR, the mean SRR rate $\lambda_{U}^{SRR}=\lambda_{U}^{SR}$.

An HR procedure takes place when a user performs a cell changing while being active. A user is considered active from the triggering of the SR procedure to the triggering of the associated SRR event. Let $P_{UA}$ be the likelihood that a user is active at a given time, and $CCR$ the mean user cell crossing rate, i.e., the mean number of cell crossings per unit time. Then the mean HR arrival rate per user ($\lambda_{U}^{HR}$) is: 

\begin{equation}{\lambda_{U}^{HR}=CCR\cdot P_{UA}}
\label{eq:lambdaHR}
\end{equation}

On the one hand, assuming that each user moves at constant speed with a random direction uniformly distributed between $[0, 2\pi)$ (fluid-flow mobility model), the $CCR$ is: 

\begin{equation}
 CCR=\frac{\overline{v} \cdot B}{\pi \cdot S}
\label{eq:ccr}
\end{equation}

 where $\overline{v}$ is the mean user speed and $B$ is the perimeter of the cell coverage area $S$. 
 
 On the other hand, to calculate $P_{UA}$ 
 we have to define the time extension of a user activity period ($T_{ua}$) as the interval from the end of an application activity period to the inactivity timer expiration or the next activity period, whichever comes first. 
 If $X$ is a generic random variable to model the elapsed time from the end of an activity period to the start of the next one, $T_{ua}$ will follow the same distribution as X, but upper truncated to the value of $T_{I}$. Thereby, the expected value of $T_{ua}$ can be computed with eq. \ref{eq:ta}:

\begin{equation}{\overline{T}_{ua}(X)=T_{I}\cdot P(X>T_{I})+\int_{0}^{T_{I}} x\cdot f_{X}(x) \, dx}
\label{eq:ta}
\end{equation}

Finally, $P_{UA}$ is $\lambda_{S}$ times the amount of time that a user is active within a session:

\begin{equation}P_{UA}=\lambda_{S} \cdot (\overline{N}\cdot \overline{T}_{on}+(\overline{N}-1)\cdot \overline{T}_{ua}(D)+\overline{T}_{ua}(T_{IS}))
\label{eq:pa}
\end{equation}

   
   

\section{Queuing Model}
\label{sec:queuing-model}

  To simulate the system architecture described in section \ref{sec:system-model}, in this work we provide a queue model based on \cite{Vilaplana2014} which considers the layout of a typical cloud processing chain. 

 Our queue model assumes that in the ingress of the computing cloud, a balancer schedules each control request to the proper NFV instance  (see Figure \ref{fig:nfvqueue}). Each NFV instance can access a shared database and response with control messages via an egress switch.
 
\subsection{Model Description}


\begin{figure}[t]
\begin{center}
\includegraphics[width=1.0\columnwidth]{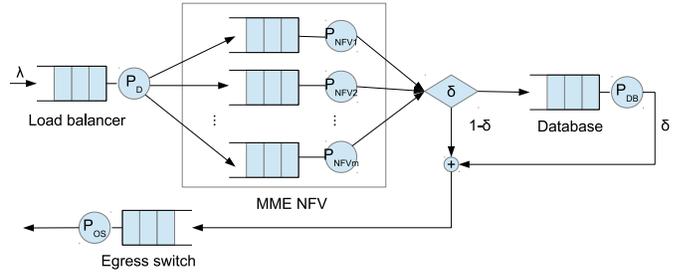}
\end{center}
\caption{Queue model of the virtualization facility.}
\label{fig:nfvqueue}
\end{figure}

 In our model, the shared database and the server which balances the control requests among the NVF processors are modeled with a single processor queue (fig. \ref{fig:nfvqueue}). The processing NVF pool is modeled by a set of queues and processors to allow the parallel processing of the control messages. 
 
 The common database is accessed during each transaction with a probability $\delta$. Since in this work we consider that every transaction requires querying the database, $\delta=1.0$.
 
  The service time of the ingress switch and database queue processors have been obtained experimentally (see section \ref{sec:cpu-numerical-results}). The egress switch service time is calculated by using the output line rate. vMME NVF service times depend on the type of message received. 
\section{Numerical Results}
\label{sec:numerical-results}
\subsection{Experimental Setup}

Our evaluation framework includes two software tools: a generator of procedure calls and a queuing system simulator.

The generator of procedure calls is implemented in the ns-3 simulator \cite{ns3}. It implements the traffic models presented in section \ref{sec:traffic-models} and the corresponding network signaling. The simulation scenario is based on the dense urban information society scenario of the METIS project \cite{metis}. It is composed of $12$ access points distributed regularly in a $4 x 3$ grid over a rectangular area of size $387m\,x\,552m$. The coverage area for each access point is rectangular with dimensions of $138m\,x\,129m$. The users move across the area following a fluid-flow mobility model.
The user speed is uniformly distributed between $0$ and $4.2m/s$.

The percentage of traffic generated for each type of application has been adjusted to meet the simulation guidelines of METIS project (see Table \ref{tab:traffic-models}). All users have an independent and constant uplink and downlink data rate of $300Mbps$ \cite{metis}. 
During the simulation, each control procedure taking place generates control messages which are dumped to a trace file. 

The queuing system simulator implements the queuing model presented in section \ref{sec:queuing-model} using the Matlab Simulink framework. 
The queuing model is fed with the traces produced by the previous tool. The load balancer has a service rate of $120000$ packets per second \cite{rightscal}.  The database service rate has been obtained by assuming that the database deployed in the Amazon Cloud is the Amazon Aurora database \cite{amazon-ec2}, which is reported to serve $100000$ transactions per second \cite{aurora-benchmark-2015}. The egress switch is a 10G Ethernet and able to serve $5000000$ packets per second. Table \ref{tab:procedure-response-time} shows the NVF processing times of the control messages used.

 

\subsection{NVF Processing Time Estimation}
\label{sec:cpu-numerical-results}

 To calculate the system delay, we need to estimate the time a NFV spends processing each control message. This value depends on the type of control procedure served. Given a CPU processing capacity, we can estimate the delay of processing a message by assessing the average number of CPU instructions required for running a particular procedure.
 
 To do this, we have considered the CPU characteristics of a real cloud service configuration from the \emph{Amazon Elastic Compute Cloud (EC2)} \cite{amazon-ec2}. Additionally, we have implemented in C the code of the functions which are invoked in the vMME for each procedure. Although our implementation may differ from real MME implementations, we think that our version executes similar tasks as the real ones. 
 
  After compiling the code, we measured the number of CPU instructions executed for every procedure by means of profiling tools. 
  Table \ref{tab:procedure-response-time} provides the delays calculated for the \emph{EC2 m3.xlarge}  virtual instance of the Amazon EC2 service \cite{amazon-ec2}. The average computing capacity of this type of instance is been measured in \cite{iosup2011performance} as $11.38\cdot 10^9$ float operations per second.
  

   



\begin{table}[tb]
\centering
{\small

\begin{tabular}{l l l}
\hline
    Procedure & Number of instructions & processing time (s)\\ 
\hline
\hline
$SR_1$ & 1.45e+06 &  12.74e-05 \\ 
$SR_2$ & 1.07e+06 & 9.40e-05 \\ 
$SRR_1$ & 1.07e+06 & 9.40e-05 \\ 
$SRR_2$ & 1.07e+06 & 9.40e-05 \\ 
$SRR_3$ & 1.06e+06 & 9.32e-05 \\ 
$HR_1$ & 1.07e+06 & 9.40e-05 \\ 
$HR_2$ & 1.07e+06 & 9.40e-05 \\ 
\hline
\end{tabular}
}
\caption{Processing times for the number of instructions measured, and their processing time for the \emph{m3.xlarge} instance.}
\label{tab:procedure-response-time}
\end{table}

\subsection{Signaling Procedures Rate}
\label{sec:arrival-rate-results}
\begin{figure}[tb]
\begin{center}
\includegraphics[width=1.0\columnwidth]{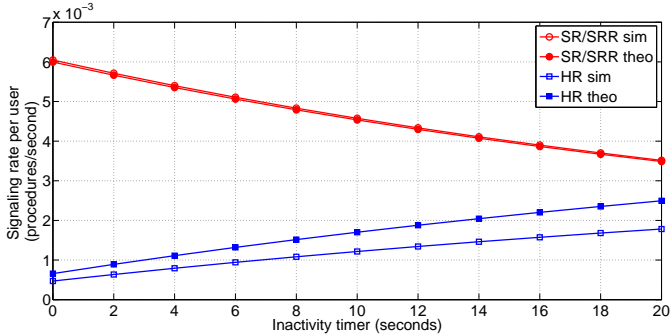}
\end{center}
\caption{Control procedures arrival rates versus user inactivity timer}
\label{fig:lambdas}
\end{figure}

To characterize the control messages arrival rate, we generated a signaling trace for 20000 users. The mean arrival rates for the different signaling procedures were computed for several $T_I$ values (see Figure \ref{fig:lambdas}). 
The results show that the SRs and SRRs rates decreases with $T_I$. That is because the higher the timer value, the smaller the probability the timer runs out within an inter activity period. Thus, the user stays 
 in the active state between consecutive application activity periods, avoiding the need for triggering procedures to reserve and release resources. 
Conversely, the HRs rate increases with the timer value, since the user remains active longer after an application activity period. Consequently, there is a higher chance that a user is active when a cell crossing event takes place.
 The root-mean-square errors between the experimental and predicted rates for SR and HR procedures ($4.07\cdot 10^{-5}$ and $5.0\cdot 10^{-4}$, respectively) demonstrate that the analytical expressions proposed are well fitted to the experimental data. 
The higher prediction error for the HR procedure rate is due to the fluid-flow mobility model implementation: a bounce-back strategy is employed when a user reaches an edge of the geographical area. That decreases the $CCR$ per user in comparison with the predicted by the fluid flow model expression.       

\subsection{System Delay}

\begin{figure}[tb]
\begin{center}
\includegraphics[width=1.0\columnwidth]{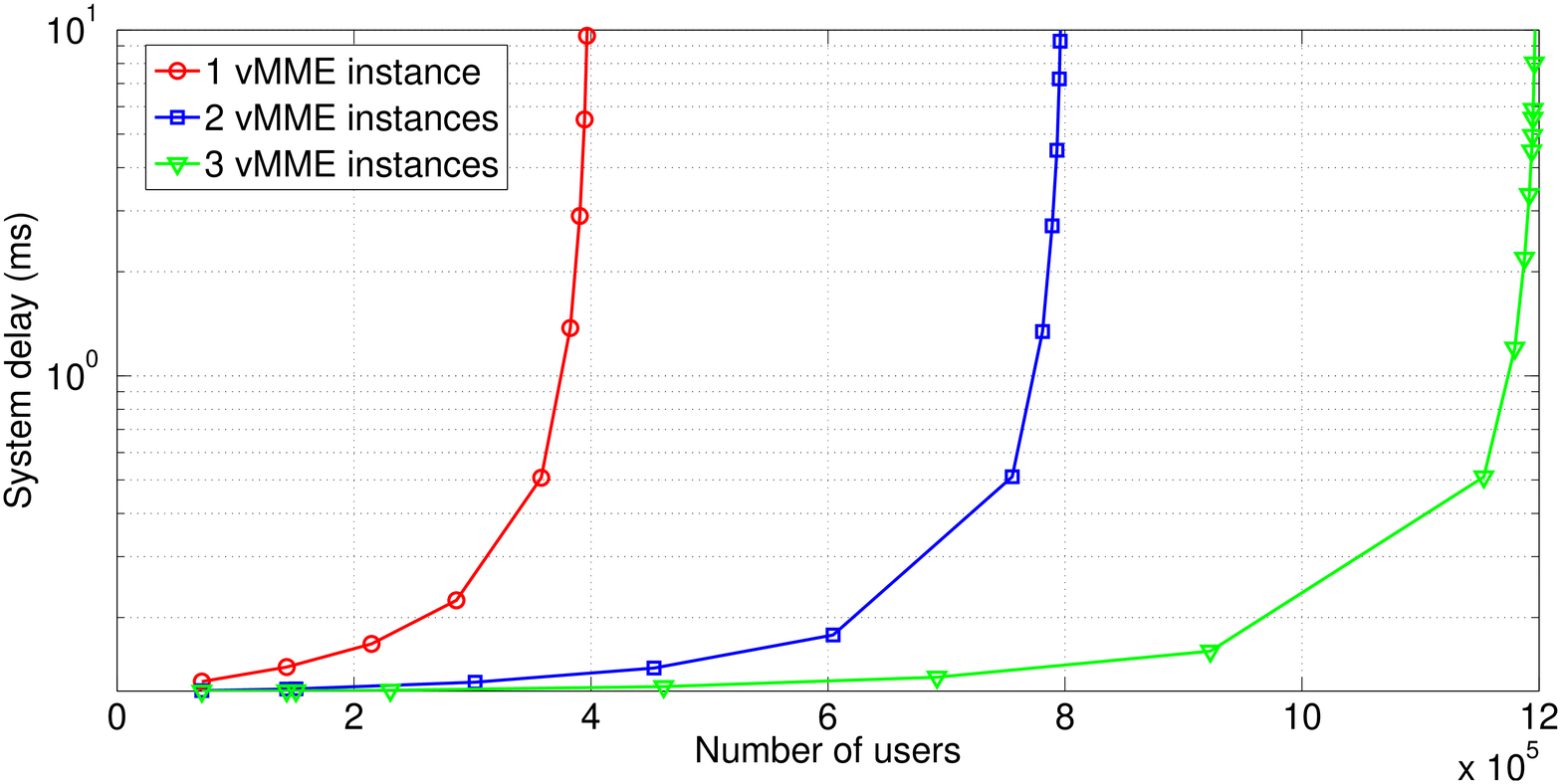}
\end{center}
\caption{Overall system delay}
\label{fig:system-delay}
\end{figure}

 Most mobile networks standards requirements define a delay budget to perform the control procedures. 
 In order to evaluate the delay of our system, we generated a signaling trace for $1200000$ users and a $T_I=10$ seconds. 
Figure \ref{fig:system-delay} depicts the system delay versus the number of users and vMME instances. The system delay grows exponentially with the number of users. There is a point where the number of vMME instances cannot withstand the control messages arrival rate and the system delay shoots up. At this point, a new MME instance must be added to cope with the control plane workload within the budget delay.

 Observing the results in Figure \ref{fig:system-delay}, we could derive a simple criterion to calculate how many vMME instances are needed to maintain the overall latency below the a given threshold in this scenario.
 
For instance, if we consider a system delay budget of 1 ms, the experimental results show that the system described is able to cope with up to $374740$, $773210$ and $1173900$ users for one, two and three vMME instances respectively. The resulting control loads are $3883$, $8016$ and $12169$ signaling procedures per second, respectively. With these results, we can predict the number of vMME instances $m$ given a number of users $u$ as $m(u)=\lceil 2.50 \cdot 10^{-6} \cdot u + 6.36\cdot 10^{-2} \rceil$.  Please note that with the traffic models parameters considered, these control workloads correspond up to $1.2\cdot 10^6$ users. Other traffic and processing times parameters may need a different equation. 
 


\section{Conclusions}
\label{sec:conclusions}
 
  In this paper we propose a queue model of a virtual MME in a datacenter, estimating its processing time for several types of control procedures.  Additionally, we have developed analytical expressions to predict the rate of UE signaling events for a given application traffic model. The accuracy of the proposed expressions has been verified by simulation.
  Using this framework we have characterized the service delay of the control signaling of a vMME which serves the traffic workloads expected in future mobile networks. 
    
  This characterization will help to design virtual resources allocation algorithms to provide, given a number of users, the desired service within the allowed delay threshold. Experimentally, we have shown that, given a processing delay threshold of 1 ms and a per user parameterized application traffic model, three vMME instances are able to cope with the signaling control traffic generated by more than 1170000 users in a datacenter with nowadays processing power.



\section*{Acknowledgment}
  
This  work  is  partially  supported  by  the  Spanish  Ministry  of  Economy   and   Competitiveness   (project   TIN2013-46223-P), FEDER and the Spanish Ministry of Education, Culture and Sport (FPU grant 13/04833).

%
\bibliographystyle{IEEEtran}
\bibliography{IEEEabrv,references}

\end{document}